# Macroscopic Coulomb Blockade Model of Density Wave Transport

John H. Miller, Jr.

*Department of Physics and Texas Center for Superconductivity*
*University of Houston*
*Houston, Texas 77204-5932*

**Abstract**

Charge and spin density waves are highly correlated electron systems that can transport an electric current. A model is discussed in which a field $E$ induces the creation, by quantum tunneling, of pairs of oppositely charged solitons and antisolitons in a density wave. Pair creation is blocked, for fields below a threshold value $E_T$, because the increase in electrostatic energy would violate energy conservation. When $E > E_T$, this Coulomb blockade mechanism time-correlates the pair creation and annihilation events. The model provides a natural explanation for the extremely small phase displacements below threshold suggested by NMR and other experiments.

**PACS numbers:**   73.23.Hk,  72.15.Nj,  71.45.Lr,  75.30.Fv,  11.25.Sq

## Macroscopic Coulomb Blockade Model of Density Wave Transport

Charge[1] and spin[2] density waves are extraordinary examples of spontaneous symmetry breaking, in which pairs of electrons and holes condense into a new ground state. A charge density wave (CDW) forms in a conducting linear chain compound when the electronic charge density becomes modulated: $\rho(x) = \rho_0(x) + \rho_1\cos[2k_Fx - \phi(x,t)]$, where $k_F$ is the Fermi wavevector and $\phi$ is the phase. A spin-density wave (SDW), which is manifested by a modulation of the spin density $\Delta S(x) = \Delta S_0 \cos[2k_Fx - \phi(x,t)]$,[2] forms as a result of electron-electron interactions and is equivalent to two out-of-phase CDWs for the spin-up and spin-down subbands. Although pinned by impurities, a DW can be induced to transport a current by applying a $dc$ field above a threshold value $E_T$.

The simplest model of DW pinning is a sine-Gordon model,[3] which represents the phase as an elastic string in a periodic pinning potential: $V(\phi) = V_0[1 - \cos\phi]$. The "vacuum states" are the minima located at multiples of $2\pi$. In the absence of dissipation, a DW would oscillate in one of these wells at the pinning frequency $\omega_0$. Finite wavelength modes, or phasons,[4] can propagate at the phason velocity, $c_0 = \mu^{-1/2}v_F$, in the short wavelength limit. The Fröhlich mass ratio,[5] $\mu = M_F/m$, is large ($\sim 10^3$) in a CDW since the electrons must drag along a moving lattice distortion, but is smaller in an SDW. The phasons are found[6] to have fermion partners in the form of dressed electrons (quantum solitons) by performing a Bose-Fermi transformation.[7] The quantum sine-Gordon model is well known to be equivalent to the massive Thirring model of interacting fermions. The bosonic form of the sine-Gordon Hamiltonian for a DW, excluding Coulomb interactions, can be written as:

$$H_0 = \int dx \left\{ \frac{\Pi^2}{2D} + \frac{1}{2} D c_0^2 \left( \frac{\partial \phi}{\partial x} \right)^2 + D \omega_0^2 [1 - \cos\phi] \right\}, \quad (1)$$

where $D = \mu \hbar / 4\pi v_F$ (per spin per chain), and the canonical momentum density is given by $\pi = D \partial \phi / \partial t$.

A topological soliton (antisoliton), of width $l_\phi = c_0/\omega_0 \sim 1$ μm $(\gg l_{dw})$ and energy $D_\phi = (2\mu^{1/2}/\pi)\hbar \omega_0 \sim 100$ μeV (per chain per spin), can be constructed[8] by advancing (reducing) the phase between vacuum states and minimizing the energy (Fig. 1(a)). This particle, which is described by a phase variation $\phi_\pm(x) = 2\pi n \pm 4\tan^{-1}[\exp(x/l_\phi)]$, is composed of dressed electrons and carries a net charge $\mp e^* = \mp(n_c/n)e$ per chain per electron spin, where $n_c/n$ is the fraction of condensed electrons. A free soliton of momentum $p$ would behave as a relativistic particle of energy $[m_\phi^2 c_0^4 + c_0^2 p^2]^{1/2}$, with a mass defined by $D_\phi = m_\phi c_0^2$. When a field $E$ is applied, a soliton-antisoliton (S-S') pair of separation $L = 2D_F/e^*E$ can be created by quantum tunneling.[8] In a related picture,[9,6] a pair is created when a quantum soliton Zener tunnels a distance $L$ through a tiny gap of energy $2D_\phi$. When the 3-D coherence due to interchain coupling is considered, this energy becomes scaled by the number ($\sim 10^6$) of parallel chains within a phase-coherent domain, as will be discussed later. The Euclidean action $S_E^{(1)}$ for producing a pair is roughly the product of energy-time uncertainties, i.e. $S_E^{(1)} \sim \Delta E \Delta t$, where $\Delta E \sim 2D_F$, and $\Delta t \sim L/c_0$. The probability of creating a pair is thus $P \sim \exp(-S_E^{(1)}/\hbar) \sim \exp(-E_0/E)$, where the Zener activation field is $E_0 \sim (2D_\phi)^2/\hbar c_0 e^*$, which is reduced in magnitude by S-S' interactions[6] and disorder.[10]

When Coulomb interactions are neglected, the above models predict a Zener-type current-field characteristic, $J \sim E \exp[-E_0/E]$, which lacks a sharp threshold field. This prediction is altered dramatically, however, by including the electrostatic energy[11] of the internal field produced by a pair. A related discussion of field-induced "quark-antiquark" pair creation in the (1+1)-D massive Schwinger model is provided by S. Coleman,[12] who points out the existence of a threshold field. Figure 1(a) shows a soliton-antisoliton pair, displaying the phase $\phi(x)$ and excess charge density $\delta\rho(x) = -(e^*/2\pi)\partial\phi/\partial x$ as functions of position. The pair, which is analogous to a parallel plate capacitor of separation $L$ and cross-sectional area $A$ per chain (per spin), produces an internal field of magnitude $E^* = e^*/\varepsilon A$. Here $\varepsilon$ is the low-frequency dielectric constant, which is enormous (~$10^8 \varepsilon_0$) in density waves and may include an intrinsic contribution $\varepsilon_{DW}$ from the "bare" DW, as well as an additional contribution $\varepsilon_s$ due to screening by normal carriers.[13,14] The electrostatic energy density is $\frac{1}{2}\varepsilon E'^2$, where $E'(x)$ is the local field. When an external field $E$ is applied, the difference between the electrostatic energy of a state with a pair and that of the vacuum is obtained by integrating over an appropriate volume:

$$\Delta U = \frac{1}{2}\varepsilon A \int dx \left[ E'^2(x) - E^2 \right] = \frac{1}{2}\varepsilon A L \left[ (E \pm E^*)^2 - E^2 \right]$$

$$= e^* L \left[ \frac{1}{2} E^* \pm E \right]. \qquad (2)$$

Note that $\Delta U$ is positive if $|E| < \frac{1}{2} E^*$, so conservation of energy forbids the vacuum to produce a pair if the field is less than a threshold value $E_T \equiv \frac{1}{2} E^*$. The threshold voltage across a region of length $L$ will thus be $e^*/2C$, where the capacitance is given by $C = \varepsilon A/L$. Such inhibition of tunneling by electrostatic interactions is known as Coulomb blockade, and has been widely studied[15] in small tunnel junctions, where the threshold voltage for single-electron tunneling is $e/2C$.

Once pairs have nucleated, the field in the regions between the soliton and antisoliton partners (represented by the colored regions in Fig. 1(b)) will be $E - e^*/\varepsilon A$, where $\varepsilon$ decreases with field[16] above threshold. Because the field is reduced, Coulomb blockade will inhibit pair creation in these colored regions[17] until all the kinks have been annihilated. The phase is thus prevented from advancing further until the phase reaches the next vacuum state (Fig. 1(b)). As a result, the S-S' nucleation and annihilation events will become correlated in time at the drift frequency $\omega_D/2\pi = I_{DW}/e^*$, where $I_{DW} = (e^*/2\pi) \partial \varphi/\partial t$ is the DW current per chain (per spin). The coherent *ac* oscillations, often observed in DWs above threshold, are somewhat analogous to those observed during time-correlated single-electron tunneling (SET) in tunnel junctions. Charge is transferred by repulsive charge solitons in linear arrays of such junctions,[18] resulting in time-correlated tunneling events and coherent *ac* oscillations with frequency *I/e*. The total current density in a DW system is given by $J(t) = \varepsilon \partial E/\partial t + (e^*/2\pi A)\partial \phi/\partial t + \sigma_n E$, where the first term is the displacement current, the second is due to tunneling of solitons according to this model, and the last term is the shunt current carried by the normal electrons. The corresponding expression for correlated SET is $I(t) = CdV/dt + edn/dt + G_s V$, where $dn/dt$ is the tunneling rate and $G_s$ is the shunt conductance. If the total current is constant, then the voltage will increase until it is abruptly reduced by a tunneling event during each cycle, resulting in coherent voltage oscillations.

One prediction of the Coulomb blockade model discussed here is that the phase displacement $\phi$ below threshold can be much smaller than would be expected classically. The classical sine-Gordon prediction is readily obtained by minimizing the energy, setting $\delta H_0/\delta \phi = 0$ (with inclusion of a term $(e^*/2\pi)Ex \partial \phi/\partial x$ describing coupling to the field). In the absence of additional electrostatic contributions, this results in a phase given by $\phi = \sin^{-1}[E/E_c]$, where $E_c \equiv D\omega_0^2/e^*$ represents the classical critical field. The phase would then vary continuously from *zero* to $\pi/2$ as the field is increased from *zero* to $E_T$ ($=E_c$), as shown in Fig. 2. Numerical simulations[19] and renormalization group calculations[20] of a classical model based on pinning by random impurities also predict substantial phase displacements below threshold.

The electrostatic energy due to the applied field $E$ and internal fields $-(\phi/2\pi)E^*$ generated by phase variations is most readily expressed by adding a term of the form:

$$H_E = \int dx \, u_E [\phi + q_E]^2,$$

(3)

to the Hamiltonian. Here, $u_E = \frac{1}{2} eA(E^*/2^1)^2$, $q_E \equiv -2^1 E/E^*$, and $f$ is measured with respect to its value at the contacts ($= 2^1 n$ in this idealized model). The total Hamiltonian, $H = H_0 + H_E$, thus becomes the bosonic form of the massive Schwinger model.[11,12] If we now set $\delta H/\delta f = 0$, as before, then the static phase displacement below threshold will satisfy $2^1 E_c \sin f + (E_T/^1)f = E$. Figure 2 shows the predicted static phase displacements as functions of applied field below threshold for several values of $E_T/E_c$. When the pinning energy $Dw_0^2$ is greater than the electrostatic energy $u_E$ (i.e. when $E_c > E_T = E^*/2$), the phase displacement will be small, and given by $f \sim E/2^1 E_c$.

There is compelling evidence that the average CDW phase displacement is indeed quite small for fields below threshold in NbSe$_3$ samples. Ross et al.[21] measured the nuclear magnetic resonance (NMR) frequency shift of NbSe$_3$ crystals at 77 K (below the upper Peierls transition) with the use of spin echos to observe the NMR signals. A *dc* pulse, equal to ¾ of the threshold field, was applied to the crystals between the 90° and 180° *rf* pulses. The measured echo integral was found to be consistent with a Gaussian distribution of CDW displacements of full width 4°, with an average displacement <*f*> of only 2°, as shown in Fig. 2. By contrast, the classical sine-Gordon model predicts a phase displacement of about 50° when $E/E_T = $ ¾.

The above NMR results are consistent with observations[16] that all *ac* properties in the small-signal limit, including the linear, direct mixing, harmonic mixing, second harmonic generation, and third harmonic generation responses, are essentially independent of bias field below threshold in orthorhombic TaS$_3$ samples. For example, Fig. 3 shows that the measured dielectric constants are independent of *dc* bias field below threshold, which is in accordance with the Coulomb blockade model if $E_T << E_c$. These experiments have also been found to be consistent with photon assisted tunneling theory[22] for bias fields above threshold. On the other hand, the classical sine-Gordon[23] and random pinning[19,20,24] models predict that the dielectric constant should increase substantially as $E$ approaches $E_T$, as shown in Fig. 3. The classical predictions are in clear contradiction with the experimentally observed behavior. The Coulomb blockade model thus not only explains the experimental observations, but also highlights some serious deficiencies of the classical paradigm.

The proposed model also explains the current-field characteristics of a variety of samples. The $J_{DW}$ -$E$ curves fit extremely well[25] with a Zener characteristic which has been modified to include the threshold field $E_T$: $J_{DW}(E) = \sigma_{max} [E - E_T] \exp[-E_0/E]$. The threshold field will be larger than $E_0$ in samples where the electrostatic energy is dominant, while the opposite will be true when the pinning energy dominates. The *I-V* curve will be rounded, showing upward curvature, when $E_0 > E_T$, but will scale roughly linearly with $E - E_T$ when $E_0 < E_T$. This latter behavior is similar to that of an ideal tunnel junction exhibiting Coulomb blockade, and is also observed in $NbSe_3$ samples with very low impurity concentrations. The limiting conductivity $\sigma_{max}$ depends on the extent to which DW motion is damped by viscous, dissipative interactions with the uncondensed, normal carriers. When $E_0 < E_T$ and $\sigma_{max}$ diverges, the differential conductance will diverge, as observed in fully gapped materials, such as blue bronze,[26] at low temperatures when the normal carriers are frozen out.

The transition temperatures of DW formation can be quite high, 215 K in $TaS_3$ for example. The interchain coupling and resulting 3-D coherence must therefore suppress thermal excitation of the S-S' pairs, since $2D_\phi^{(1)} << k_B T$ for a single chain. If the phase is correlated across $N \sim 10^6$ parallel chains, then the energy of a collective $S^{(N)}$-$S'^{(N)}$ pair will be $2D_\phi^{(N)} = 2N D_\phi^{(1)} >> k_B T$, so the thermal excitation probability, $\exp[-2N\Delta_\phi^{(1)}/k_B T]$, will be suppressed. However, the tunneling probability will be unaffected, as can be seen by examining the parameters that determine the Zener activation field $E_0^{(1)} \sim (2D_\phi^{(1)})^2/\hbar\, c_0 e^*$. If we treat the tunneling events involving *N* parallel chains (or, in *k-space*, *N* transverse wavevectors $k_\perp$) as being statistically correlated, then the effective charge $e^{*(N)} = Ne^*$ and energy $E_k^{(N)} = N E_k^{(1)} = [(N\Delta_\phi^{(1)})^2 + (N\hbar c_0 k)^2]^{1/2}$ of a soliton will be scaled up, so the gap $D_\phi^{(N)} = N\Delta_\phi^{(1)}$ and limiting slope $(\hbar c_0)^{(N)} = N\hbar c_0$ of the energy dispersion relation will each be scaled up by $N$. The activation field thus becomes $E_0^{(N)} \sim (2N\Delta_\phi^{(1)})^2/[(N\hbar c_0)(Ne^*)] \sim E_0^{(1)}$.

The above scaling argument suggests that Planck's constant should be scaled by $N$ when using the total Euclidean action, $S_E^{(N)} = NS_E^{(1)}$, to compute the probability of nucleating an $N$-chain soliton pair, i.e. $\exp(-S_E^{(N)}/N\hbar) = \exp(-S_E^{(1)}/\hbar)$. In order to justify this argument, let us note that the momentum of an $N$-chain quantum soliton is $P^{(N)} = Np^{(1)} = N\hbar k$, which indicates that either $\hbar$ or $k$ ought to be scaled by $N$. If the quantum solitons (dressed electrons) became tightly bound in *real space*, then the *wavevector* $k = 2^1/\lambda_{\text{de Broglie}}$ would be scaled up. However, the DW electrons, although interacting, remain delocalized throughout the crystal, suggesting that $\hbar$ should be scaled up in this case. The situation here may be analogous to Josephson tunneling of Cooper pairs through an insulating barrier, where the tunneling amplitude is that of a single pair even though the $N$ pairs form a condensate. A complete theoretical treatment would require replacing a single path (or functional) integral in Euclidean space by many individual functional integrals for the (thermally inaccessible) degrees of freedom represented by $N$ coupled scalar fields $f_n$.

The model discussed here suggests that DW depinning is a *high temperature* collective quantum phenomenon that can occur above room temperature in some materials, such as $NbS_3$. This model is consistent with a body of evidence that not only supports the quantum depinning hypothesis, but also refutes[27] the predictions of the classical deformable models. The implications of a quantum mechanism of DW depinning are potentially profound and far-reaching. The proposed scaling of Planck's constant by the number of quantum degrees of freedom, for example, could have an impact on inflationary models of the universe, which propose that the universe originated from a quantum nucleation event at the beginning of time. It should also be noted that a long Josesphson junction (JJ) is described by a sine-Gordon Hamiltonian and may be the dual of a DW. Indeed, a theory of quantum nucleation of soliton-antisoliton (vortex- antivortex) pairs has been proposed[28] to explain the rounded $V$ vs. $I$ curves observed in long JJs. Thus, by utilizing charge-flux duality[29] in reduced dimensional systems, some aspects of the theory proposed here may apply to tunneling of flux vortices[30] in superconductors and long JJs.

The author thanks C. Salmeron, J. Claycomb, and M. Bell for their assistance, and acknowledges helpful conversations with M. Paczuski, C. Ordoñez, E. Prodan, W. P. Su, and C. S. Ting. This work was supported, in part, by the State of Texas through the Texas Center for Superconductivity at the University of Houston and the Texas Higher Education Coordinating Board Advanced Research Program, and by the Robert A. Welch Foundation.

[9] John Bardeen, *Phys.Rev.Lett*. **42**, 1498 (1979); *ibid* **45**, 1978 (1980); *ibid* **55**, 1002 (1985).

[10] I. V. Krive and A. S. Rozhavsky, *Phys. Lett.* A**132**, 363 (1988).

[11] I. V. Krive and A. S. Rozhavsky, *Solid State Commun.* **55**, 691 (1985).

[12] S. Coleman, *Annals of Phys.* **101**, 239 (1976).

[13] J. R. Tucker, W. G. Lyons, J. H. Miller, Jr., R. E. Thorne, and J. W. Lyding, *Phys. Rev.* B **34**, 9038 (1986).

[14] Following J. M. Ziman, *Principles of the Theory of Solids,* (Cambridge, 1972), the intrinsic dielectric constant is approximately $e_{DW}/e_0 \sim 1 + \omega_p^2/\omega_0^2$, where the plasma frequency is given by $\omega_p^2 \sim n_c e^2/\varepsilon_0 m$. The contribution due to screening by normal carriers is about $e_s/e_0 \sim 1 + (k_F l)^2 [1-n_c/n]$, where l is the length over which DW phase variations create excess charge. This length is roughly that of a phase-coherent domain, which decreases with impurity concentration.

[15] D. V. Averin and K. K. Likharev, in *Mesoscopic Phenomena in Solids*, Ed. by B. L. Altshuler, P. A. Lee, and R. A. Webb (North-Holland, Amsterdam, 1991).

[16] J. H. Miller, Jr., R. E. Thorne, W. G. Lyons, J. R. Tucker, and John Bardeen, *Phys. Rev.* B**31**, 5229 (1985); J. H. Miller, Jr., Ph.D. thesis, University of Illinois at Urbana-Illinois, 1986 (unpublished).

[17] Actually, additional pairs *can* nucleate in such regions if the field exceeds $3e^*/2e(E)A$, $5e^*/2e(E)A$, etc. and the pairs don't annihilate too quickly. (The various "critical fields" would be determined self-consistently, since $e(E)$ decreases with field above threshold.) Such decay channels may cause additional kinks in the field-dependent conductivities, similar to a "Coulomb staircase," in some samples.

**FIGURE CAPTIONS**

**Fig. 1.** (a) A materialized soliton-antisoliton pair, showing the position-dependent phase $f(x)$, excess charge density $dr(x) = -(e^*/2^1)\_f/x$, and internal field $E^* = e^*/eA$.

(b) The time-evolution of the phase $f(x,t)$, illustrating the nucleation and subsequent annihilation of three pairs during the first cycle. The field in the colored regions is $E - E^*$. Additional pair creation in these regions is blocked by Coulomb blockade until the pairs have annihilated and the phase has advanced to its next vacuum state.

**Fig. 2.** Predicted phase displacements below threshold, based on the classical sine-Gordon model (solid line) and Coulomb blockade model for several values of $E_T/E_c$. The results obtained by Ross et al.[21] are indicated by the symbol.

**Fig. 3.** Predicted field-dependent dielectric constant according to the classical overdamped sine-Gordon model,[23] the classical random pinning models,[19,20,24] where $|f| \equiv |1-E/E_T|$, and the Coulomb blockade (C.B.) model ($E < E_T$). The predicted dielectric constants for $E > E_T$ were obtained using photon-assisted tunneling (P.A.T.) theory.[22] The experimental dielectric constants were determined by measuring the imaginary part of the *ac* admittance, which peaked at 200 MHz, of a $TaS_3$ sample at 185 K.[16]

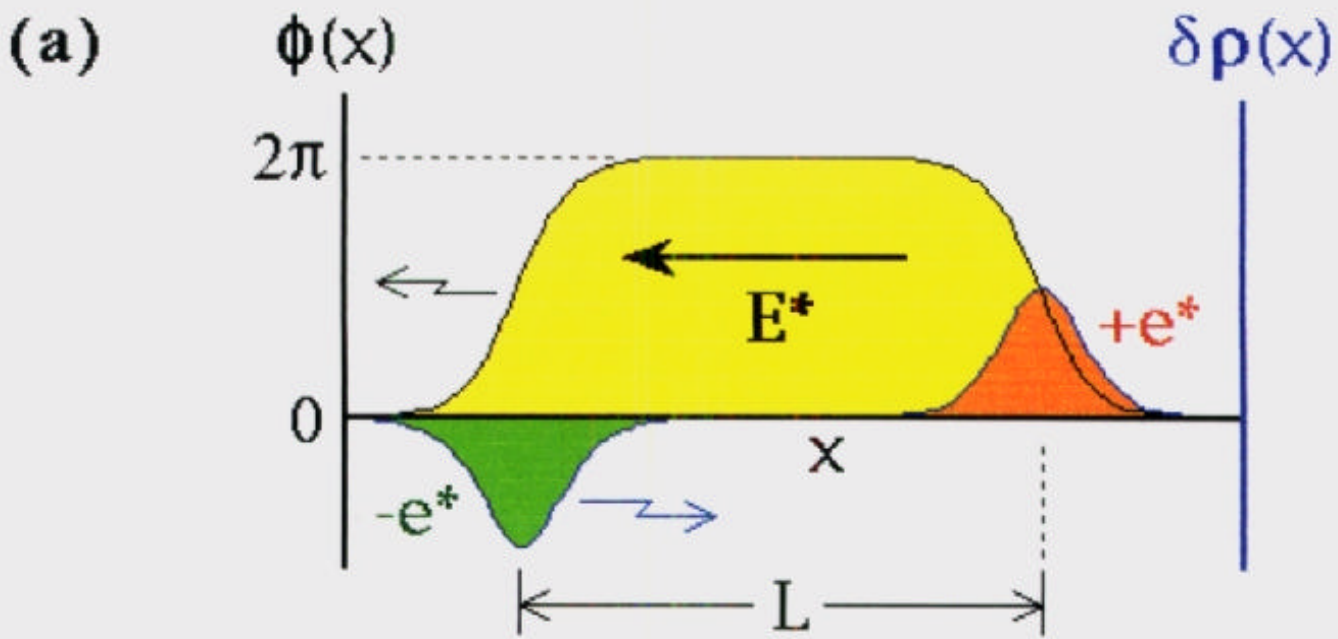

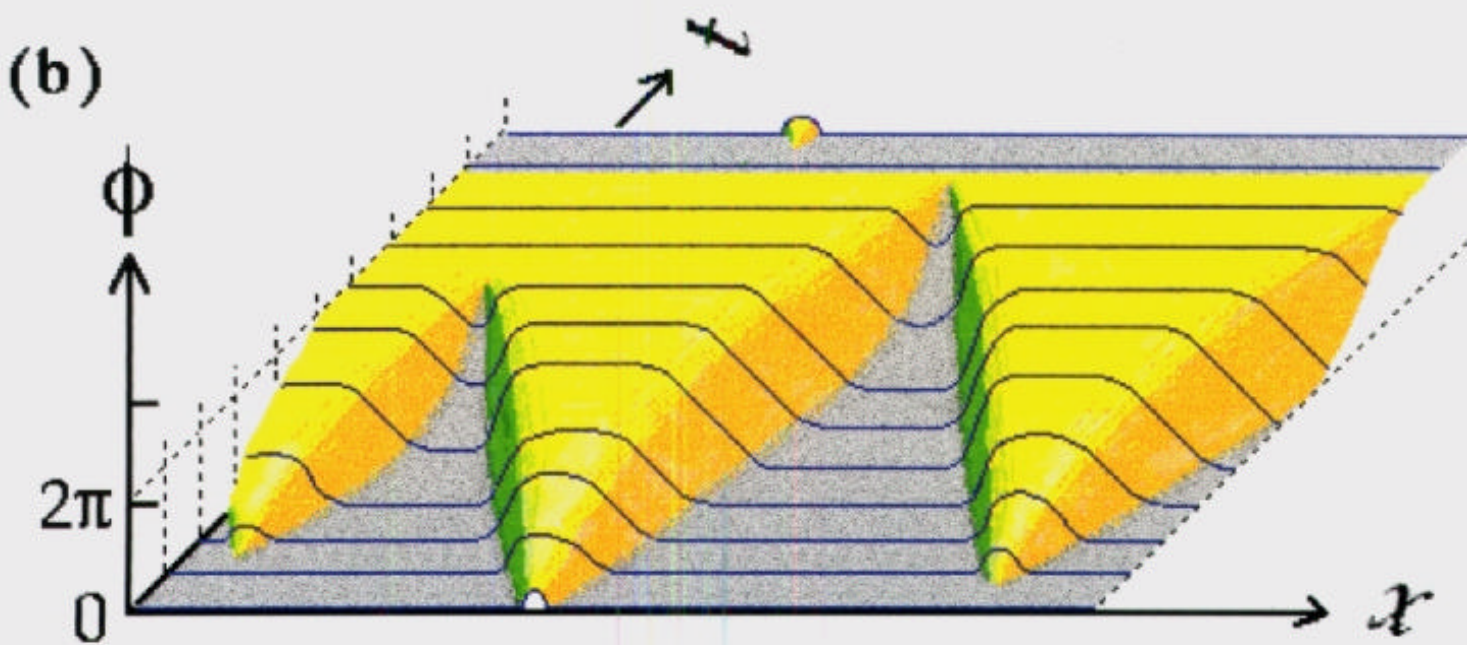

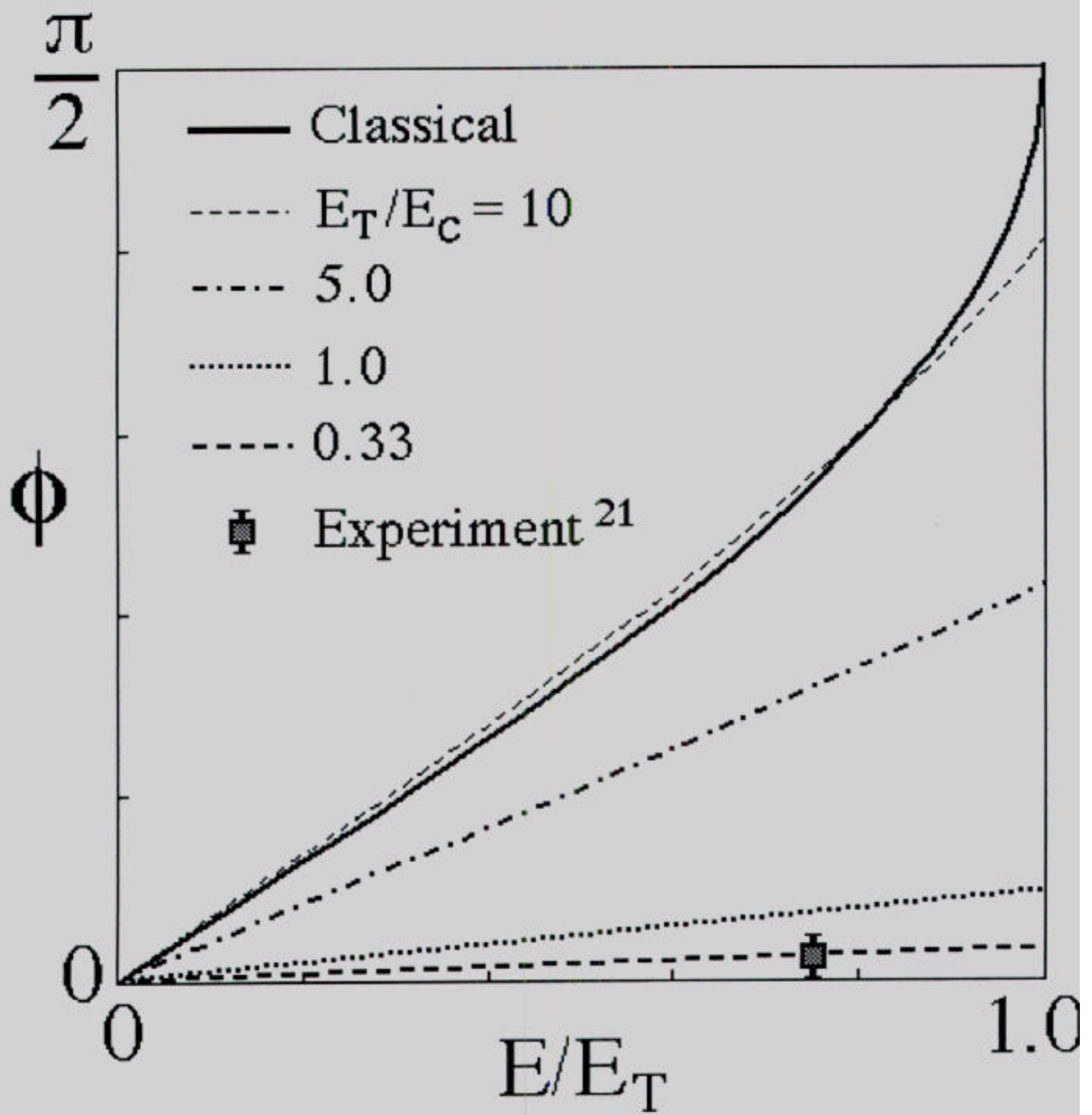

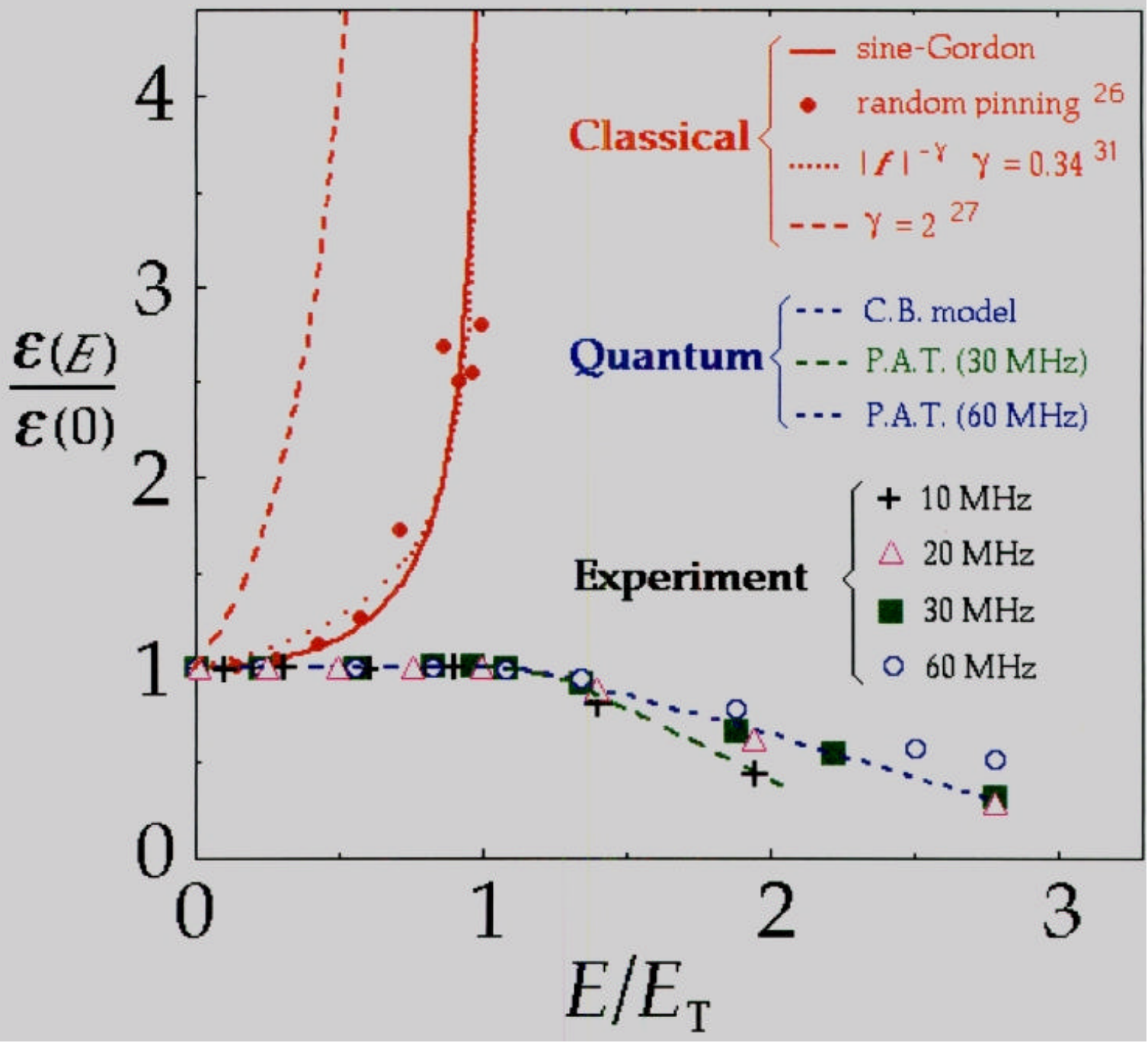